\documentclass[12pt]{article}
\def\beq{\begin{equation}}
\def\eeq{\end{equation}}
\def\bea{\begin{eqnarray}}
\def\eea{\end{eqnarray}}
\def\bef{\begin{figure}}
\def\enf{\end{figure}}
\def\S{{\bf S}}
\def\C{{\bf C}}
\def\Z{{\bf Z}}

\def\P{{\bf P}}

\def\RP{{\bf RP}}
\def\CC{{\cal C}}
\def\CF{{\cal F}}

\def\CN{{\cal N}}
\def\CO{{\cal O}}

\def\We{{W_{eff}}}

\def\ba{\begin{array}}
\def\ea{\end{array}}
\def\bce{\begin{center}}
\def\ece{\end{center}}

\def\ev#1{\langle#1\rangle}

\def\pa{\partial}

\def\IC{{\relax\hbox{$\inbar\kern-.3em{\rm C}$}}}
\def\ID{\relax{\rm I\kern-.18em D}}
\def\IE{\relax{\rm I\kern-.18em E}}
\def\IF{\relax{\rm I\kern-.18em F}}
\def\IG{\relax\hbox{$\inbar\kern-.3em{\rm G}$}}
\def\IGa{\relax\hbox{${\rm I}\kern-.18em\Gamma$}}
\def\IH{\relax{\rm I\kern-.18em H}}
\def\II{\relax{\rm I\kern-.18em I}}
\def\IK{\relax{\rm I\kern-.18em K}}

\def\IQ{\relax\hbox{$\inbar\kern-.3em{\rm Q}$}}

\begin{document}
\begin{titlepage}
\rightline{HUTP-01/A016} \rightline{HU-EP-01/14}
\rightline{hep-th/0104037}
\def\today{\ifcase\month\or
January\or February\or March\or April\or May\or June\or July\or
August\or September\or October\or November\or December\fi,
\number\year} \vskip 1cm \centerline{\Large \bf Orientifold,
Geometric Transition and} \vskip 1mm \centerline{\Large \bf Large
N Duality for SO/Sp Gauge Theories} \vskip 1cm \centerline{\sc
Jos\'e D. Edelstein$^{a,}$\footnote{edels@lorentz.harvard.edu},
Kyungho Oh$^{a,}$\footnote{On leave from Dept. of Mathematics,
University of Missouri-St. Louis, oh@hamilton.harvard.edu} and
Radu Tatar$^{b,}$\footnote{tatar@physik.hu-berlin.de}} \vskip 1cm
\centerline{{ \it $^a$ Lyman Laboratory of Physics, Harvard
University, Cambridge, MA 02138, USA}} \vskip 1mm \centerline{{\it
$^c$ Insitut fur Physik, Humboldt University, Berlin, 10115,
Germany}} \vskip 2cm \centerline{\sc Abstract} \vskip 0.2in

We extend the large $N$ duality of four dimensional $\CN = 1$ supersymmetric
Yang--Mills theory with additional chiral fields and arbitrary
superpotential recently proposed by Cachazo, Intriligator and Vafa to the
case of $SO/Sp$ gauge groups. By orientifolding the geometric transition, we
investigate a large $N$ duality between $\CN = 1$, $SO/Sp$ supersymmetric
theories with arbitrary superpotential and an Abelian $\CN = 2$ theory with
supersymmetry broken to $\CN = 1$ by electric and magnetic
Fayet--Iliopoulos terms.

\vskip 1in \leftline{March 2001}
\end{titlepage}
\newpage

\section{Introduction}

String theories on conifolds have been used to study large $N$ gauge
theories. In \cite{kw}, it was conjectured that there is a large $N$
duality  between type IIB string theory on ${\bf AdS_5 \times
T^{1,1}}$ and $\CN = 1$ conformal supersymmetric theory in four
dimensions with gauge group $SU(N) \times SU(N)$. The dualities
have been soon extended to more general conifold type
singularities \cite{mp}. The Renormalization Group flow has been
studied after fractional D--branes are introduced which corresponds
to relevant deformation of the conformal field theory~\cite{kn,ot2}.
Furthermore, it was shown \cite{ot3} that the moduli space
becomes non-commutative if discrete torsion is present on the
conifold which gives rise to marginal deformation.

Some months ago, a new direction has been developed concerning type II
strings at conifold singularities, and a large $N$ duality has been
proposed by Vafa between gauge systems with $\CN=1$ supersymmetry in four
dimensions and superstrings propagating on non-compact Calabi--Yau
manifolds with RR fluxes \cite{vafa}. Via geometric transition, type IIA
theory with $N$ D6 branes wrapped on a special Lagrangian 3-cycle of a
Calabi--Yau is dual to type IIA theory on K\"ahler deformed Calabi--Yau
where the 3-cycle has been replaced by an exceptional $\P^1$. On
the K\"ahler deformed side, the D--branes disappear and are replaced
by $N$ units of RR fluxes through the exceptional $\P^1$. This duality
proposal is based on the embedding of the Chern--Simons/topological
string duality of \cite{gova} into superstring theory. From the M--theory
perspective, this duality can be obtained by a geometrical flop on 7
dimensional manifolds with $G_2$ holonomy \cite{achar,amv}. The
supergravity picture of this geometrical transition was then studied
in \cite{en}.

More recently, this large $N$ duality proposal has been extended to four
dimensional $\CN =1$ supersymmetric $SU(N)$ Yang--Mills theory with an
adjoint chiral field and arbitrary superpotential by Cachazo,
Intriligator and Vafa \cite{civ}. This theory is claimed to be dual,
through a similar geometric transition, to an Abelian theory where
supersymmetry is broken to $\CN=1$ by means of Fayet--Iliopoulos terms.
Most strikingly, the glueball chiral superfields,
whose first components are the gaugino bilinears, are identified with
certain compact periods in a non-compact Calabi--Yau, while the cutoff of
the otherwise divergent non-compact periods gives an elegant explanation
of the running of the coupling constant. Furthermore, the duality in
\cite{civ} allows to compute the exact quantum effective superpotential of
the low energy $SU(N)$ theory.

In this paper, we extend the results of \cite{civ} to the case of $SO/Sp$
gauge groups. The main idea is to obtain a large $N$ duality for the
$SO/Sp$ gauge groups by acting by orientifolding on both sides of the
duality proposed for $SU(N)$ gauge theories. The complex conjugation
provides an orientifolding on the conifold which can be extended to both
K\"ahler and complex deformed conifolds to be considered. In the case
without superpotential, the three cycles $\S^3$ on the complex deformed
conifold will be invariant under the orientifolding and in the dual
picture, the fluxes will be through an $\RP^2$ cycle instead of the
$\P^1$ of $U(N)$, after the $Z_2$ projection. The $\CN = 1$, $SO/Sp$
super Yang--Mills theory with adjoint chiral field $\Phi$ and arbitrary
tree level superpotential can be geometrically engineered by perturbing a
conifold to a non-compact Calabi--Yau space which has only conifold
singularities, and then orientifolding. In this process, the compatibility
with the orientifolding will impose a constraint on complex and K\"ahler
deformations of the Calabi--Yau space implying a particular arrangement
of the 3-cycles and the 2-cycles along a complex line. In the K\"ahler
deformed Calabi--Yau, an $\RP^2$ will be located at the origin of the
complex line while pairs of $\P^1$ cycles will be located along its
imaginary axis symmetrically with respect to the origin. The orientifolding
will map one $\P^1$ into the other in the pair. In \cite{sv}, the same
orientifolding process has been considered in the context of a duality
between $SO/Sp$ Chern--Simons gauge theory and topological string theory
on an orientifold of the small resolution of the conifold.

Based on the orientifolded geometric transition explained above, we propose
a large $N$ duality for $\CN = 1$, $SO/Sp$ supersymmetric theories with
arbitrary superpotential and an Abelian $\CN = 2$ theory with supersymmetry
broken to $\CN = 1$ by electric and magnetic Fayet--Iliopoulos terms. The
$SO/Sp$ group is broken in a generic $\CN = 1$ vacuum, and the theory
reduces at low energies to an Abelian $U(1)^n$ supersymmetric theory with
$N-n$ massless monopoles that admits a Seiberg--Witten formulation. An
important check on the proposed duality is to show that the periods
obtained from the reduced Seiberg--Witten theory arising in the
world--volume of the D5 branes on the K\"ahler deformed conifold are
consistent with the ones obtained by using the symplectic geometry on the
complex deformed conifold. This implies that the couplings of the Abelian
gauge fields agree in both dual theories. To show this, we have used the
results of \cite{aotdec} following the ideas of \cite{civ}. We show that
the equations of both hyperelliptic curves of degree $4n+2$ agree up to
a degree $2n$ starting from the highest degree.

The results obtained could be connected to the ones constructed
via brane configurations with D4 branes, NS branes and orientifold
four or six planes where one rotates the NS branes with respect to
each other \cite{hoo}. The rotation is similar to giving mass or
introducing more general superpotentials for the adjoint chiral field.

The content of the paper is as follows. In section 2 we
geometrically engineer the general $\CN = 1$, $SO/Sp$ theory with
an adjoint chiral field and a generic superpotential. We give a
detailed discussion of the required orientifolding, and compute
the effective superpotential showing that the regularization of
non-compact periods give raise to the expected running of the
coupling constant, with the appropriate $\beta$ function, for
$SO/Sp$ gauge theories. We consider the action by orientifolding
on the geometric transition of \cite{civ}, and obtain the
geometric realization of symmetry breaking patterns of the form
$SO(N) \to SO(N_0) \times U(N_1) \times \cdots \times U(N_n)$, or
$Sp(N) \to Sp(N_0) \times U(N_1) \times \cdots \times U(N_n)$. In
section 3 we investigate strongly coupled field theory implications from
the proposed large $N$ duality for $SO/Sp$. In particular, we compare
the reduced Seiberg--Witten curve with one obtained from the dual geometry.
In passing, we give a new derivation of the exact low energy effective
superpotential induced by generic microscopic deformations within the
framework of the Whitham hierarchy formalism of softly broken $\CN=2$
supersymmetric gauge theories.

\section{Geometric Engineering of SO(N) gauge theories with
superpotential}

We begin by explaining the geometric backgrounds we will be dealing
with in this paper.  The conifold is a three dimensional hypersurface
singularity in $\C^4$ defined by: \bea \label{fold} {\cal C}:
\quad x^2 + y^2 +z^2 + w^2 = 0 ~. \eea We consider two kinds of
deformations of the conifold, {\em i.e.} complex and K\"ahler. A
complex deformed conifold given by \bea \label{comdef} k := x^2 +
y^2 + z^2 + w^2 -\mu =0 \eea will be considered. This is
isomorphic to $T^*\S^3$ as a symplectic manifold after the
rotation of symplectic structure by the phase of $\mu$. On the
other hand, the K\"ahler deformed conifold to be considered will
be a small resolution of the conifold. By this process, the
singular point of the conifold $\CC$ will be replaced by a $\P^1$
with normal bundle $\CO (-1) + \CO(-1)$.

In type IIA string theory, a four dimensional $\CN = 1$, $U(N)$
supersymmetric gauge theory is obtained by wrapping $N$ D6 branes
on the $\S^3$ in the complex deformed conifold. In \cite{vafa},
Vafa has proposed a duality, in the large $N$ limit, between this
theory and type IIA superstrings without $D$--branes propagating
on the K\"ahler deformed conifold. This duality emerges as the
embedding of the large $N$ Chern--Simons/topological string
duality of Gopakumar and Vafa \cite{gova} in ordinary
superstrings. The branes are replaced by $N$ units of RR flux
through $\P^1$, and also NS flux on the dual four cycle. Acting by
orientifolding on the duality of \cite{gova}, the large $N$
duality of $SO$ and $Sp$ Chern--Simons gauge theory on $\S^3$ and
topological strings on an orientifold of the small resolution of
the conifold has been studied in \cite{sv}. This last paper also
discuss the embedding of this duality in ordinary superstrings.
Note that the orientifolding acts on the conifold $\CC$ via the
complex conjugation \bea \label{conj1} (x,y,z,w) \to (\bar{x},
\bar{y}, \bar{z}, \bar{w}) ~, \eea and can be extended both to complex
and K\"ahler deformations considered above provided that $\mu$ is
real. On the complex deformed conifold $T^*\S^3$, the special
Lagrangian $\S^3$ is invariant under this complex conjugation and
hence the orientifold is a O6 plane wrapping it \cite{sv,gomis}.
To see the effect of the orientifolding on the K\"ahler deformed
conifold, we introduce new coordinates \bea a = x + i y ~, ~~b = z
+ i w ~, ~~c= x- i y ~, ~~d = -z + i w ~, \eea in terms of which
the singular conifold is given by \bea ac -bd = 0 ~. \eea The
complex conjugation (\ref{conj1}) corresponds to the action \bea
\label{conj2} a \to \bar{c} ~, ~~b \to -\bar{d} ~, ~~c \to \bar{a}
~, ~~d \to -\bar{b} ~. \eea The K\"ahler deformation considered
above can be described as a union of two smooth 3-dimensional
complex manifolds, namely $M_1 = \{(a',b,c,d) \in \C^4 | d =
a'c\}$ and $M_2 = \{(a,b',c,d) \in \C^4 | c = b'd\}$ glued
together by the natural identification and the extra condition $b=
a b'$. The smooth manifold obtained in this way maps onto the
conifold. Notice that the singular point $(0,0,0,0)$ of the
conifold is now replaced by $\P^1$ whose coordinates are given by
$a'$ on $M_1$ and $b'$ on $M_2$. One can see that the complex
conjugation sends $a'$ to $-\bar{b'}$ which is the antipodal map
on $\S^2$. The quotient by this action will be ${\bf RP}^2$.

The $\CN = 1$ supersymmetric gauge theory on the world--volume of
$N$ D6 branes sitting on top of O6 planes wrapping the $\S^3$ of
the complex deformed conifold, has $SO(N)$ or $Sp(N)$ gauge group
depending on the choice of the sign for the world--sheets with
crosscaps. By the arguments of \cite{vafa}, this theory is
equivalent to type IIA string theory on the orientifold of the
K\"ahler deformed conifold \cite{sv}. We may give a mirror
description of this large $N$ duality which reverses the direction
of the transition. Now, in the large $N$ limit, $SO(N)$ or $Sp(N)$
theories obtained from type IIB theory by wrapping $N$ D5 branes
on the $\P^1$ in the orientifold of the K\"ahler deformation of
the conifold, are equivalent to type IIB theory on the
corresponding orientifold of the complex deformed conifold
(\ref{comdef}), with $N$ units of RR flux through $\S^3$. There is
also some $NS$ flux through the non-compact cycle. The value of
$\mu$ is fixed by the fluxes and this is captured by a
superpotential for the chiral field $S$, whose first component is
proportional to $\mu$ \cite{vafa}. We may assume that $\mu$ is
real after rotating it back by its phase. A rational three form
\bea \Omega = \frac{dx \, dy \,dz}{\partial k/ \partial w} \sim
\frac{dx \, dy \,dz}{\sqrt{\mu - x^2 -y^2 -z^2}} ~, \eea will be a
holomorphic form on the deformed conifold (\ref{comdef}). On the
complex deformed conifold, there is a single compact cycle $A
\cong \S^3$ and the corresponding dual non-compact cycle $B$. The
$A$ period of the holomorphic 3-form $\Omega$ is $S$. There are
$N$ units of RR flux though $A$, and also NS flux $\alpha$ through
$B$; $\alpha$ is identified with the bare coupling of 4d
$SO/Sp$ gauge theory similarly to what happens for $SU$
\cite{vafa}. To compute the periods, consider a projection $p$
from the conifold $\CC$ to $x$-plane. Then the inverse image of
the real interval $[-\sqrt{\mu},\sqrt{\mu}]$ is the 3-cycle $A$
and that of the line segment $[\sqrt{\mu}, \infty)$ will be the
3-cycle $B$. Thus, the $A$-period is given by \bea \label{s} S =
\int_A \Omega = \int_{[-\sqrt{\mu}, \sqrt{\mu}]} \int_{p^{-1}(x)}
\Omega = \frac{1}{2\pi i}\int_{-\sqrt{\mu}}^{\sqrt{\mu}} dx
\sqrt{x^2 -\mu} = \frac{\mu}{4} ~, \eea as in the case of $SU(N)$
\cite{civ}. The $B$ period is divergent, and thus a cutoff
$\Lambda$ must be introduced as
\begin{eqnarray}
\label{pi} \Pi & = & \int_B^{(\Lambda)} \Omega =
\int_{[\sqrt{\mu}, \Lambda^{3/2}]} \int_{p^{-1}(x)} \Omega =
\frac{1}{2\pi i} \int_{\sqrt{\mu}}^{\Lambda^{3/2}} dx \sqrt{x^2
-\mu} \nonumber \\ & = & ~\frac{1}{2\pi i} \left( \frac{1}{2}
\Lambda^3 - 3 S \log \Lambda - S (1 - \log S) \right) +
O(1/\Lambda) ~.
\end{eqnarray}

The effect of the fluxes in the four dimensional theory is an
effective superpotential of the form \cite{tv}
\bea \label{potgeom} \We = \int
\Omega \wedge (H^{RR} + \tau N^{NS}) ~, \eea where $\tau$ is the
complexified coupling constant of type IIB strings. If we wrap $N$
D6 branes on $S^3$, the effective action after orientifolding can
be written as \bea \We = \left( \frac{N}{2} \mp 1 \right) \Pi -
2\pi i \alpha S ~, \eea because only half of the $N$ units of RR
flux are left by the orientifolding operation and the orientifold
carry RR charge of $\mp 4$ units for, respectively, $SO$ and $Sp$.
There is an additional $\pm \Pi$ coming from the $\RP^2$
worldsheet of unoriented string.
As in \cite{civ}, inserting (\ref{s}) and (\ref{pi}) into
(\ref{potgeom}) and neglecting an irrelevant constant, we have
\bea \We = \left( \frac{N}{2} \mp 1 \right) \left( 3 S \log
\Lambda + S (1 - \log S) \right) - 2\pi i \alpha S ~, \eea so
finite results are reached by regularizing $\alpha$ in such a way
that the following combination is a constant \bea 3 \left(
\frac{N}{2} \mp 1 \right) \log \Lambda - 2\pi i \alpha = const.
\eea Now it is clear that, if we identify $\alpha$ with the bare
coupling of $SO$ or $Sp$ gauge theories as $2\pi i\alpha =
8\pi^2/g^2(\Lambda)$, the previous equation reproduces the running
of the coupling constant \bea \frac{8\pi^2}{g^2(\Lambda)} = 3
\left( \frac{N}{2} \mp 1 \right) \log \Lambda + const. ~, \eea
with the appropriate $\beta$ function, provided we also identify
$\Lambda$ with the scale of the theory. It is also immediate,
following the ideas of \cite{civ}, to show that there are $N \mp
2$ supersymmetric vacua where $S \neq 0$ for appropriate normalization
of the superpotential. Following \cite{vafa}, $S$ is
identified with the $SO/Sp$ glueball chiral superfield.

In Calabi--Yau string vacua, there is a hypermultiplet which
becomes light at conifold singularities in the moduli space and
can be excited \cite{strom}. Turning on fluxes causes
supersymmetry to softly break from $\CN = 2$ to $\CN = 1$. The
$\CN = 2$ vector multiplet consists of an $\CN = 1$ chiral
superfield $S$ and an $\CN = 1$ $U(1)$ photon. As orientifolding
reverses the sign of $U(1)$, the $U(1)$ gauge symmetry is broken
down to $\Z_2$. This $\Z_2$ symmetry is to be identified with the
global $\Z_2$ symmetry in the $\CN = 1$ $SO(N)$ gauge theory. On
the deformed conifold, the gauge theory on $N$ D6 branes wrapping
$\S^3$ is $\CN = 1$ $U(N)$ gauge theory. By orientifolding, the
$U(N)$ adjoint representations become $O(N)$ representations, but
as the $O(N)$ group is disconnected, the gauge theory becomes that
of $SO(N)$ with $O(N)/SO(N) = \Z_2$ as a global symmetry. In the
case of $SO(2N)$, there are particles created by the Pfaffian
$\mbox{Pf} = (1/N!) \epsilon^{i_1j_1 ...i_Nj_N} \Phi^{i_1j_1}
\cdots \Phi^{i_N j_N}$ of the adjoint field $\Phi$ that are
charged under $\Z_2$ and then annihilate in pairs \cite{witt}.
Since the Pfaffian operator vanishes for $SO(2N+1)$ and $Sp(N)$,
we do not expect corresponding particle states in such cases.

We shall now concentrate on the case of $\CN =1$ $SO(N)$ gauge
theory with matter $\Phi$ in $(\frac{1}{2}N (N +1) -1)$
dimensional traceless tensor representation of $SO(N)$, {\em i.e.}
adjoint representation, with superpotential \bea \label{treesup}
W_{tree} (\Phi) = \sum_{k=1}^{n+1}\frac{g_{k}}{2k}\mbox{Tr}
~\Phi^{2k} ~. \eea
The geometric engineering is essentially the same as in $\cite{civ}$.
We only need to take care of orientifolding. When $W_{tree}(\Phi)=0$,
the field theory is $\CN = 2$ Yang--Mills theory. The geometric
background will be the blow--up of
the singular locus given by $y=z=w=0$ on the hypersurface defined
by \bea y^2 + z^2 + w^2 = 0 ~. \eea Then we have a one dimensional
family of $\P^1$'s along $x$-axis. Each $\P^1$ has normal bundle
$\CO(-2) + \CO(0)$. From the geometry, it is clear that we can
identify the blow--up of the above hypersurface with the total space
of the normal bundle of $\P^1$ and $x$-axis can be identified with
$\CO(0)$ direction. If we wrap $N$ D5 branes on a $\P^1$ in the
family, we obtain an $\CN =2$ $U(N)$ gauge theory on their
uncompactified world--volume. The one dimensional moduli of $\P^1$
corresponds to the moduli of the Coulomb branch in $\CN =2$ gauge
theory through identification of the position of each brane with
an eigenvalue of the adjoint chiral field $\Phi$. On this geometric
background, we consider the orientifolding on $\CO(-2) + \CO(0)$
over $\P^1$ induced by the complex conjugation \bea x \to
{\bar{x}} ~, ~~y \to {\bar{y}} ~, ~~z \to \bar{z} ~, ~~w \to
\bar{w} ~, \eea on the ambient space $\C^4$. Because of the
orientifolding we are considering, the freedom to move the D5
branes is constrained. Since the eigenvalues of $SO(N)$ adjoint
field appear in pairs $ia, -ia$ (where $a$ is a real number), the
D5 branes can be moved only in pairs along the imaginary $x$-axis.
Here we have chosen the complex form of $SO$ adjoint so that the
eigenvalues will be purely imaginary numbers which is consistent
with our choice of the orientifolding. The pair is in the
reflection of each other under the orientifolding. In order to
make perturbation by the superpotential (\ref{treesup}), we need
to freeze the $\P^1$'s with the normal bundle $\CO (-2) + \CO(0)$
at the particular locations of $x$ given by the zeros of \bea
W'(x) = g_{n+1} x \prod_{j=1}^n (x^2 + a_j^2) ~, \quad a_j >0 ~.
\eea To explain the geometry of $\CO (-2) + \CO (0)$ over $\P^1$,
we introduce two copies of $\C^3$ with coordinates $z, x, u$
(resp. $z',x',u'$) for the first (resp. second) $\C^3$. Then
$\CO(-2) + \CO (0)$ over $\P^1$ is obtained by gluing two copies
of $\C^3$ with the identification: \bea \label{-2+0} z' =
\frac{1}{z} ~, ~\quad x = x' ~, ~\quad u' = u z^2 ~. \eea The $z$
(resp. $z'$) is a coordinate of $\P^1$ in the first (resp. second)
$\C^3$. Similarly, $x,x'$ (resp. $u,u'$) are coordinates of
$\CO(0)$ (resp. $\CO(-2)$) direction. Now we perturb this geometry
by the following change in (\ref{-2+0}): \bea \label{def-2+0} z' =
\frac{1}{z} ~, ~\quad  x =x' ~, ~\quad u' =u z^2 + W'(x) z ~, \eea
and in this way we now obtain $\P^1$'s only where $W'(x) = 0$,
{\em i.e.} $x = 0$ and $x = \pm i a_j$, whose normal bundle is $\CO(-2)+
\CO(0)$.  Unlike the family of $\P^1$'s before the perturbation,
we cannot move these $\P^1$'s. Under the orientifolding, the $\P^1$
located at $x=0$ becomes $\RP^2$ and the $\P^1$ located at $i a_j$ will
map to the $\P^1$ located at $-i a_j$. From now on, we will assume
that $a_j$'s are distinct and nonzero. We will consider the cases of
$SO(2N)$ and $SO(2N+1)$, the $Sp(N)$ case following straightforwardly.
We can distribute the $2N$ (resp. $2N+1$) D5 branes among the vacua
$x = 0$ and $x = \pm i a_j$ by wrapping $2N_0$ (res. $2N_0+1$) branes
around $\RP^2$ at $x = 0$ and $N_j$ branes around $\P^1$ at $\pm i a_j$,
where $\sum_{j=0}^{n} N_j = N$. For the former, $\RP^2$ is stuck on the
orientifold and it gives $\CN=1$, $O(2N_0)$ (resp. $O(2N_0+1)$) gauge
theory on the world--volume. Otherwise, the orientifold projection
identifies the states at $i a_j$ with those of $- i a_j$ and we
instead obtain $U(N_j)$ gauge theory. So this is a geometric
realization of the breaking of \bea O(N) \to O(N_0) \times
U(N_1) \times \ldots \times U(N_{n}) ~. \eea To present the large $N$
duality, we blow--down these $\P^1$'s to a singular hypersurface
in $\C^4$ which has been described in \cite{civ}. After changing
variables, the blown--down Calabi--Yau is given by \bea
\label{blown-down} W'(x)^2 + y^2 + z^2 + w^2 = 0 ~. \eea In our case,
$W'(x)$ having distinct roots, we have $2n+1$ distinct conifold
singularities. If we take a small resolution at every singular point,
then we obtain a set of $\P^1$'s whose normal bundle is $\CO(-1) +
\CO(-1)$. These $\P^1$'s are called exceptional. Under the
orientifolding, the exceptional $\P^1$'s behave the same way as before.

Following the large $N$ duality proposal of \cite{civ}, we consider a
complex deformation of (\ref{blown-down}). Topologically, this process
creates $n+1$ finite size $\S^3$'s which can be thought of as been
shrunken to zero size at the singularities of (\ref{blown-down}). At
each conifold point of (\ref{blown-down}),
there is a one dimensional complex deformation space parameterized
by a complex number $\mu_i$. As we have $2n +1$ conifold singular
points for a generic $W'(x)$, the complex deformation space will
be $2n+1$-dimensional. Being constrained by the orientifolding, the
actual deformation space is $n+1$ dimensional and parametrized by
$\mu_0, \ldots , \mu_n$. This can be achieved by a polynomial $f(x)$
of degree $2n$ in $x$ with values $f(0) = \mu_0$ and $f(\pm i a_j) =
\mu_j$. In fact, $f$ is a function of $x^2$. Hence, the complex
deformation is given by
\bea \label{compd} g := W'(x)^2 + f(x) + y^2 + z^2 + w^2 = 0 ~.
\eea Now the exceptional $\P^1$'s have been replaced by the finite
size $\S^3$'s. Under orientifolding, the 3-sphere $\S^3$ located
at $x=0$ is invariant,
 and the 3-sphere $\S^3$ located at $x=i a_j$ maps
to one located at $x = -i a_j$ and vice-versa.  Because of this,
we may restrict our discussion to the singular points lying over
the upper half $x$-plane, {\em i.e.} $x=0$ and $x_j=ia_j$. Under
the complex deformation (\ref{compd}), each of the $n+1$ critical
points in the upper half $x$-plane will split into two critical
points denoted by $x=
0^{+},0^{-},ia_1^{+},ia_1^{-},\ldots,ia_n^{+},ia_n^{-}$. On this
deformed Calabi--Yau, we can find a symplectic basis for the
3-cycles which consists of $2n+1$ compact  $A_i$ cycles and $2n+1$
noncompact $B_i$ cycles. We will only consider the $n+1$ 3-cycles
supported over upper half $x$-plane. As in the case of the
conifold, the rational three form \bea \Omega = 2 \frac{dx \wedge
dy \wedge dz}{{\partial g}/{\partial w}} \eea will be a
holomorphic three form on this deformed Calabi--Yau. The periods of
$\Omega$ over $A_j$ cycles supported on the upper half
$x$-plane are given by
\bea
S_0 = \pm \frac{1}{2\pi i} \int_{0^-}^{0^+} \omega ~~~~~~~
S_j = \pm \frac{1}{2\pi i} \int_{ia_j^-}^{ia_j^+} \omega ~, \eea
where the sign depends on the orientation; the periods over the dual
$B_j$ cycles are
\bea
\Pi_0 = \frac{1}{2\pi i} \int_{0^+}^{\Lambda_0^{3/2}} \omega ~~~~~~~
\Pi_j = \frac{1}{2\pi i} \int_{ia_j^+}^{\Lambda_0^{3/2}} \omega ~.
\eea Here $\omega$ is obtained by integrating $\Omega$ over the
fibers of the projection to $x$-coordinate as before. Thus \bea
\omega = dx ~(W'(x)^2 + f(x))^{1/2} ~. \eea
Under such a projection, the complex deformed conifold maps into a
hyperelliptic curve of the form
\bea
\label{cur}
y^2 = W'(x)^2 + f(x) ~.
\eea
After the transition, the D--branes give rise to $N_j$ units of $H_{RR}$
flux through the $j$-th $\S^3$ cycle $A_j$ and $\alpha$ units of $H_{NS}$
flux through each of the dual non-compact $B_j$ cycles, with $2\pi i
\alpha = 8\pi^2/g_0^2$~ given in terms of the bare coupling constant
$g_0$ of the original 4d $O(N)$ field theory. Instead, the $H_{RR}$ flux
through the $\S^3$ seated at the origin is reduced by the orientifolding
procedure. We now have the superpotential
in terms of the $A_i$ and $B_i$ periods as \bea
-\frac{1}{2\pi i} \We = \left( \frac{N_0}{2} \mp 1 \right) \Pi_0 +
\sum_{j=1}^n N_j \Pi_j  + \alpha \left( \sum_{j=0}^n S_j \right) ~.
\eea

The above K\"ahler deformed Calabi--Yau   is almost identical to
that of a partial resolution of the orbifolded conifold under
$\Z_n$ action studied in \cite{ot2}. The studied partial
resolution of the orbifolded conifold by a $\Z_n$ has $n$ ordinary
conifold singularities along $n-1$ $\P^1$'s and the distance
between them is controlled by the size of  $\P^1$ cycles. The
intersection matrix of these $\P^1$ cycles is the same as the
Cartan matrix of $SU(n)$ group. In \cite{ot2}, this issue has been
studied for the duality between type IIB string theory on ${\bf
AdS_5 \times X_5}$ and the $\CN =1$ field theory living on the
world--volume of the D3 branes placed on the orbifolded conifold
singularity. Here ${\bf X_5}$ is the horizon of the orbifolded
conifold. It would be interesting to formulate a large $N$ duality
via geometric transition in this context.

\section{Field Theory Results and Large N Duality}

Based on Chern--Simons/topological string duality \cite{gova}, in
\cite{vafa} a new duality has been formulated, stating that the large
$N$ limit of the field theory obtained on $N$ D6 branes wrapping the
special Lagrangian 3-cycle of a deformed conifold is dual to type IIA
strings propagating on the blow--up of the conifold, the latter being a
$\CO (-1) + \CO (-1)$ bundle over $\P^1$. In the dual/string picture we
have $N$ units of RR flux through the $\P^1$ and NS flux through the dual
4-cycle. The decoupling limit occurs when $N$ is large and by duality
the size of the blown--up sphere gets identified with the glueball chiral
superfield and the expectation value of its lowest component corresponds
to gaugino condensation.

By mirror symmetry, the type IIB picture is obtained, where the
original $U(N)$ theory takes place on the world--volume D5 branes wrapped
on the blown--up cycle of the resolved conifold, and the dual picture is
type IIB on the deformed conifold background with fluxes but without
branes . In presence of a generic superpotential, the number of blown--up
cycles increases. The results of \cite{strom} state that the gauge group
obtained after the compactification of type IIB string is $U(1)^{n}$ were
$n$ is the number of compact 3-cycles. The $\CN = 2$ supersymmetry is
broken to $\CN = 1$ by the presence of electric and magnetic fluxes
\cite{tv} so that the field theory obtained in the deformed conifold
side is $\CN = 1$, $U(1)^{n}$ gauge theory.

The main claim of this paper is that the above $U(1)^{n}$ gauge fields
coincide with those of the theory $O(N_0) \times U(N_1) \times \ldots
\times U(N_{n})$, that results from the low energy excitations around a
generic Higgs vacuum of the $O(N)$ theory with superpotential
(\ref{treesup}), after the $SO(N_0), SU(N_1), \ldots, SU(N_{n})$ factors
get a mass gap and confine, the $Z_2 = O(N_0)/SO(N_0)$ being a global
group. The exact quantum effective coupling constants should coincide
with the above $\tau_{ij}$. In particular, the arguments of \cite{civ}
showing that $\tau_{ij}$ is the period matrix of a hyperelliptic curve
extends to our case, the curve being precisely (\ref{cur}). We do this
by using the complex deformation of the geometry from the previous
section and show that the coupling constants can be obtained from two
identified hyperelliptic curves: one is the Seiberg--Witten curve of
the field theory leaving on the world--volume of the D5 branes on the
K\"ahler deformed conifold, and the other is connected to the complex
deformation, and it is given by (\ref{cur}).

\subsection{Undeformed theory}

Before going into the details of the deformed model, we first discuss
the pure $\CN = 1$, $SO$ field theories. For the $U(N)$ case, the large
$N$ duality was formulated in \cite{civ} in purely gauge theoretic
terms. In the complex deformed side, the gluino condensation left only
the $U(1)$ subgroup. In the K\"ahler deformed side, this $U(1)$ group
was identified with the $U(1)$ gauge theory coming from the
compactification on conifold and by turning electric and magnetic
Fayet--Iliopoulos superpotential terms which softly break $\CN = 2$
to $\CN = 1$ \cite{tv,strom}.

In the presence of the orientifold, the actual situation in the
complex deformed conifold side is that on the D6 branes wrapped on
the 3-cycle we have  the group $O(N) = SO(N) \times Z_2$ where the
group $Z_2$ is a discrete one \cite{witt}. On the other hand, the
orientifold does not affect the 3-cycle but implies a $Z_2$
projection on all the other coordinates so the above
compactification on the conifold will give a $Z_2$ group instead
of $U(1)$. This geometrical $Z_2$ group is identified with the
group $Z_2$ on the world-volume theory for the D6 branes. As in
\cite{strom}, there is also a charged hypermultiplet under the
$Z_2$ group which is to be identified with the baryon field of the
$SO(N)$ theory as discussed in \cite{witt}.

Therefore the duality can be formulated in purely gauge theoretic
terms. We shall see next how things go in the presence of the
superpotential.

\subsection{Moduli Space for the Deformed Theory}

We now consider an $\CN = 1$ supersymmetric gauge theory with adjoint chiral
superfield $\Phi$ and the tree level superpotential \bea W_{tree}
= \sum_{k=1}^{n+1} \frac{g_k}{2k} \mbox{Tr} (\Phi^{2k})
\label{delw} \eea Without the superpotential, the theory would
have $\CN = 2$ supersymmetry.

The classical theory with the superpotential (\ref{delw}) has many
vacua as discussed in \cite{ls}. If we rotate the field $\Phi$ into a
$2 \times 2$ block form ${\rm diag}(x_1i\sigma_2,\ldots,x_{N}i\sigma_2)$
for $SO(2 N)$ and ${\rm diag}(x_1i\sigma_2, \ldots,x_{N}i\sigma_2,0)$
for $SO(2N+1)$, with $\sigma_2$ the Pauli matrix, the supersymmetric
ground states are given by the zeroes of
\begin{equation}
\label{moduli} W'(x) =g_{n+1} ~x \prod_{j=1}^n (x^2 + a_j^2) = 0 ~,
\end{equation}
and the ground states are labeled by a set of integers
$(N_0,N_1,\ldots,N_{n})$; each $N_j$ giving the number of
eigenvalues $x_i$ of $\ev{\Phi}$ which are equal to $0$ or $i a_j$. In
this vacuum the gauge group is broken to
\begin{equation}
\label{bre1} SO(2 N +1) \rightarrow SO(2 N_0+1) \times
U(N_1)\times\ldots\times U(N_{n})
\end{equation}
or
\begin{equation}
\label{bre2} SO(2 N)\rightarrow SO(2 N_0)\times
U(N_1)\times\ldots\times U(N_{n})
\end{equation}
with $\sum_{j=0}^{n} N_j=N$. These breaking patterns were geometrically
engineered in the previous section.

At low energies, for each $U(N_i)$ factor in (\ref{bre1})--(\ref{bre2}),
the corresponding $SU(N_i)$ develops a mass gap and confine, the remaining
$U(1)$ staying massless. In total, the low energy theory has then $U(1)^n$
symmetry, and we will identify the corresponding couplings $\tau_{ij}$
with the second derivative of the prepotential of the complex deformed
conifold.

\subsection{The Low Energy Superpotential}

In this section we would like to present a novel derivation of the
exact low--energy counterpart of the superpotential (\ref{delw}), which
is usually claimed to be given simply by its vacuum expectation
value, without further modifications to the Seiberg--Witten
solution. In principle, it seems that this is the case only for
small values (with respect to appropriate powers of $\Lambda$) of
the parameters $g_{k}$. However, it was already argued by
Seiberg and Witten in the case of deformations of $SU(2)$ super
Yang--Mills theory parameterized solely by $g_1$, that the result
is exact for any value of $g_1$ \cite{sw}.

The most natural framework to generalize this statement for a generic
deformation $\{g_k\}$ is provided by the Whitham hierarchy associated
to $\CN = 2$ supersymmetric gauge theories \cite{gkmmm,gmmm,emm}. Let us
proceed for simplicity in the case of $SU(N)$. Consider the parameters
$g_{k}$ as components of $\CN = 2$ vector multiplets $T_k$,
\begin{equation}
T_k = t_k + \tilde\theta^2 g_{k} + \cdots
\end{equation}
where the dots amount to the remaining components of the superfield. The
fields $T_k$ enters into the effective prepotential almost on the same
footing than the coordinates $A_i$ of the Coulomb branch. The main
difference is that they are monodromy invariants and so must be their
duals $T_k^D$. Moreover, notice that a semiclassical contribution to the
prepotential of the form
\begin{equation}
\delta\CF_k = T_k ~U_{k+1} ~,
\end{equation}
where $U_{k+1}$ is the chiral superfield corresponding to $\mbox{Tr}
(\Phi^{k+1})$, leads, after integration in $\tilde\theta^2$, to a
superpotential of the form
\begin{equation}
\label{supcl} \delta W_{class} = g_{k} \frac{\partial \CF}{\partial
T_k} = g_{k} U_{k+1} ~.
\end{equation}
The theory is nevertheless still $\CN = 2$ invariant since it is
constructed out of $\CN = 2$ vector multiplets. The supersymmetry
is softly broken to $\CN = 1$ if we freeze the $T_k$ fields
as\footnote{A complete discussion of the soft breaking to ${\cal
N}=1$ by means of the spurion mechanism in the case corresponding
to a quadratic superpotential was given in \cite{lura}.}
\begin{equation}
T_1 = 1/g_0^2 + \tilde\theta^2 g_1 ~~~~~~~T_k=\tilde\theta^2 g_k
~~~~ k>1 ~,
\end{equation}
$g_0$ being the bare coupling constant. It is immediate to see
that the semiclassical contribution to the effective
superpotential is (\ref{supcl}), and correspond to the microscopic
deformation we are interested in. In order to write down the low
energy effective action involving all quantum corrections, we
should be able to compute $\frac{\partial \CF}{\partial T_k}$
exactly. This is indeed possible, provided we interpret the $T_k$
superfields as Whitham slow times \cite{emm}. The Whitham hierarchy
provides a framework in which first and second order derivatives of
the prepotential with respect to the slow times can be computed
\cite{gmmm,emm}. When these times are promoted to spurion superfields,
after freezing their components either to softly break to ${\cal N}=0$
or to ${\cal N}=1$, only these lower order derivatives contribute to
the effective action. This allows writing an exact effective
potential. The case of a generic soft breaking to ${\cal N}=0$ was
addressed in \cite{emm}. When supersymmetry is broken to ${\cal
N}=1$, as in the present paper, the only non vanishing
contribution to the effective potential is given by the first
order derivative of the prepotential with respect to $T_k$, which
is exactly\footnote{The discussion here is oversimplified. The
actual situation is that these derivatives give a homogeneous
combination of the Casimirs of order $k$, but we can always
reacommodate the microscopic superpotential analogously, and call
$g_{k}$ the corresponding coefficients.} $U_{k}$. This result also
holds in the case of $SO/Sp$ gauge groups \cite{egrmm}. This shows
that the exact low energy effective superpotential (besides the
contribution of BPS massless states) is given by
\bea
W_{low} = \sum_{k=1}^{n+1} g_{k}
U_{k} ~. \eea
In the case of $SO(N)$, we should replace $U_k \to U_{2k}$.

\subsection{Strong Coupling Dynamics}

The $\CN =2$ theory deformed by (\ref{delw}) only has unbroken
supersymmetry on submanifolds of the Coulomb branch, where there are
additional massless fields besides the $u_r$. They are nothing but the
magnetic monopoles or dyons which become massless on some particular
submanifolds $\ev{u_p}$ where the Seiberg--Witten curve degenerates.
Near a point with $l$ massless monopoles, the superpotential is
\begin{equation}
W = \sqrt{2} \sum_{i=1}^{l} M_i A_i \tilde M_i + \sum_{k=1}^{n+1}
g_{k} U_{2k} ~,
\end{equation}
where $A_i$ denote the chiral superfield of the $U(1)$ vector multiplet
corresponding to an $\CN = 2$ dyon hypermultiplet $M_i$. The vevs of the
lowest components of $A_i,M_i,U_{2k}$ are written as $a_i,m_{i},u_{2k}$.
The supersymmetric vacua are at those $\ev{u_{2k}}$ satisfying:
\begin{equation}
\label{vacua}
a_i(\ev{u_{2k}}) = 0 ~, ~~~~~~~ g_k + \sqrt{2} \sum_{i=1}^{l}\frac{\pa
a_i}{\pa u_{2k}}(\ev{u_{2k}}) m_i \tilde m_i = 0 ~,
\end{equation}
for $k=1,\ldots,n+1$. The value of the superpotential at this vacuum is
given by
\begin{equation}
\We=\sum _{k=1}^{n+1}g_k \ev{u_{2k}} ~.
\end{equation}
Before orientifolding, there are $N$ massless photons in the original
$\CN = 2$, $U(N)$ theory while there are $2n+1$ massless
photons in the vacuum after the gauge group is broken as \bea
\label{vac-ori} U(N) \to U(N_0) \times U(N_1)_- \times U(N_1)_+
\times \cdots \times U(N_n)_- \times U(N_n)_+ \eea where $N_0 + 2
\sum_{i=1}^n N_i = N$. Here $U(N_j)_-$ (resp. $U(N_j)_+$) is associated
with $N_j$ D5 branes wrapping $\P^1$ located at $-ia_j$ (resp. $+ia_j$).

If we now act by orientifolding, the original $\CN = 2$, $SO$ theory
contains $N$ massless photons whereas, in the low energy theory at the
vacua (\ref{bre1})--(\ref{bre2}), only $n$ photons remain massless.
Thus, there are $N - n$ mutually local magnetic monopoles at the
orientifold vacua (\ref{bre1})--(\ref{bre2}) which become massless
and get an expectation value as in (\ref{vacua}), $\ev{m_i \tilde m_i}
\neq 0$ for $i= 1, \ldots , N-n$. The vacuum obtained from integrating
out $u_{2k}$ as in (\ref{vacua}), will give $\ev{u_{2k}}$ in terms of the
$g_k$. Solving for the supersymmetric vacua as in (\ref{vacua}), is
equivalent to minimizing $W_{low}$, subject to the constraint that
$\ev{u_p}$ lie on the codimension $N-n$ subspace of the Coulomb branch
where at least $N-n$ mutually local monopoles or dyons are massless.

Let us consider, for definiteness, the case of $SO(2N)$ gauge theory. The
corresponding discussion for $SO(2N+1)$ and $Sp(N)$ follows immediately.
Recall that the Seiberg--Witten curve for the $SO(2N)$ theory\footnote{The
Seiberg--Witten curve for $SO/Sp$ theories can be written both as a
hyperelliptic curve of genus $r$ ($r$ the rank of the group) \cite{as1}
or, instead, as a hyperelliptic curve of genus $2r-1$ with ${\bf Z_2}$
symmetry \cite{so}. The second choice is more naturally connected
to the geometrical picture where we have the ${\bf Z_2}$ symmetry.} is
\bea \label{swso}
y^2  = P_{2N}(x^2;u_{2k})^2 - \Lambda^{4N-4} x^4 ~,
\eea
where
\bea P_{2N}(x^2)= \det (x - \Phi) =\prod_{i=1}^{N} (x^2 - x_i^2)
= \sum_{i=0}^{N} s_{2i} x^{2(N-i)} ~, \eea
The $s_{2k}$ and
$u_{2k}$ are related each other by Newton's formula
\begin{equation}
2k s_{2k}+ \sum_{i=1}^{k} s_{2k-2i} u_{2i} =0, \;\;\; k=1, 2,
\cdots, N \label{newton}
\end{equation}
with $s_0=1$. This curve has $\Z_2$ symmetry. By taking quotient
of the original hyperelliptic curve (\ref{swso}) of genus $2N-1$,
we will get a hyperelliptic curve of genus $N$ which will
correctly produce the periods (and hence the field theory). The
original curve (\ref{swso}) is a double covering of the curve
obtained by the $\Z_2$ quotient. In our geometric engineering
point of view, the $\Z_2$ symmetry corresponds to the
orientifolding which had put the constraint on the deformation.
Hence we may compare $\CN =1$ theory from our large $N$ duality
with $\CN=2$ Seiberg--Witten theory in both ways, {\em i.e.} $U(2N)$
theory with orientifold symmetry with the original curve of genus
$2N-1$ with $\Z_2$ symmetry or $O(2N)$ theory obtained after
orientifolding with the curve of genus $N$ after the $\Z_2$
quotient. But since we have arranged the adjoint $\Phi$ to have
purely imaginary eigenvalues in the orientifolding, the
orientifold will produce $\Z_2$ symmetry of the curve after
changing its complex structure. Therefore, the condition on the
original curve (\ref{swso}) for having $N-n$ mutually local massless
magnetic monopoles is that, after shifting $x$ if necessary
\begin{equation}
\label{curve} y^2 = P_{2N}(x^2;\ev{u_{2k}})^2 - \Lambda^{4N-4} x^4 =
x^2 (H_{2 N - 2 n - 2}(x))^2 F_{4 n + 2}(x)
\end{equation}
where $H_{2N-2n-2}$ is a polynomial in $x$ of degree $2N-2n-2$ and
$F_{4n+2}$ is a polynomial in $x$ of degree $4n+2$. Actually, both
$H$ and $F$ are polynomials in $x^2$.

The photons which are left massless in (\ref{moduli}) have gauge
couplings which are given by the period matrix of the reduced
curve
\begin{equation}
\label{red-curve}
y^2 =F_{4n+2} (x^2;\ev{u_{2k}}) = F_{4 n + 2}(x^2; g_k,\Lambda) ~,
\end{equation}
and from this we can read the gauge couplings of the $U(1)^{n}$ group
in terms of $g_k,\Lambda$.

Let us briefly discuss about the Seiberg--Witten curve after the
quotient to clarify its connection with the number of massless
dyons at the vacuum $\ev{u_{2k}}$. The original curve (\ref{swso})
will become a curve of genus $N$ while the quotient of the reduced
curve (\ref{red-curve}) by $\Z_2$ will be of the
form \bea y^2 = x \tilde{F}_{2n+1} (x)\eea and its genus is $n$.
Here $\tilde{F}_{2n+1}(x)$ is obtained by replacing $x^2$ by $x$
in $F_{4 n + 2}(x^2)$. Hence $N-n$ cycles has shrunken and these
correspond to massless magnetic monopoles.

\noindent
Recall from the previous section that the dual Calabi--Yau geometry
is determined by
\bea
W'(x)^2 + f(x) + y^2 +z^2 +w^2 = 0 ~.
\eea
this determining the coupling constants $\tau_{ij}$,
\bea \label{tauij}
\tau_{ij} = \frac {\partial \Pi_i }{\partial S_j} ~, \eea 
that, as shown in \cite{civ} for the $SU$ case, can be identified with
the period matrix of the hyperelliptic curve (\ref{cur}). Thus, in order
to show that the $\tau_{ij}$ from (\ref{red-curve})
agrees with that obtained from the dual geometry, we need to prove
that $F_{4 n + 2}(x^2)$ is related to the superpotential by
\begin{equation}
\label{nec} g_{n+1}^2 F_{4 n + 2}(x) = W'(x)^2 + f(x) ~.
\end{equation}
Here the factor $g_{n+1}^2$ is inserted to match the coefficient
of the highest term in both sides. This identity would provide very
strong evidence for the proposed duality. It is a non-trivial statement
which is difficult to prove in general. In the $SU$ case, for example,
it was shown only for the case of cubic superpotentials. We will prove
it up to the coefficients of $f(x)$, that is, for the terms of the
polynomials of degree greater than $2n$.

To prove the statement, we denote the roots of $H_{2(l-1)}$ and
$F_{4N-4l+2}$ of the curve (\ref{curve}) with $l$ massless dyons by
$\pm p_i$ and $\pm q_i$ and so we have
\bea H_{2(l-1)}(x) = \prod_{i=1}^{l-1} (x^2 - p_i^2) ~, ~~~~~
F_{4N-4l+2}(x) = \prod_{i=1}^{2N-2l+1} (x^2 - q_i^2) ~.
\label{HyF}
\eea
We may assume that $p_i$ and $q_i$ are non-zero numbers because $x$
is factored out in (\ref{curve})\footnote{This assumption was not spelled
out explicitly in (2.17) of \cite{aotdec}.}. Then, there is a relation
between the parameters $g_{k}$ and the dyon vevs $m_i^2$ ~\cite{aotdec}
\bea g_{k} = \sum_{i=1}^{l-1}\sum_{j=0}^N s_{2(j-k)} p_i^{2(N-j)} 
\omega_i \eea
where
\bea \omega_i =
\frac{m_{i,dy}^2}{2\prod_{s\neq i}^l(p_i^2 - p_s^2) \prod_{t\neq
i}^{2N -2l}(p_i^2 - q_t^2)^{1/2}} \eea Following the arguments of
\cite{aotdec}, we obtain a convenient form for the superpotential
\begin{eqnarray}
W_{cl}'(x) & = & \sum_{k=1}^{N} g_{k} x^{2k-1} \nonumber \\
& = &\sum_{k=1}^{N} \sum_{i=1}^{l-1} \sum_{j=0}^N x^{2k-1} s_{2(j-k)}
p_i^{2(N-j)}\omega_i \nonumber \\ & = & \sum_{k=-\infty}^{N}
\sum_{i=1}^{l-1} \sum_{j=0}^N x^{2k-1} s_{2(j-k)} p_i^{2(N-j)} \omega_i \pm
\Lambda ^{2 N-2} \sum_{i=1}^l p_i^2 \omega_i  x^{-1} + {\cal O}(x^{-2})
\nonumber \\ & = & \sum_{i=1}^{l-1} \frac{P_{2N}(x^2;\ev{u_{2k}})}{x
(x^2-p_i^2)} ~\omega_i \pm \Lambda ^{2 N-2} \sum_{i=1}^l p_i^2 \omega_i
x^{-1} + \CO (x^{-2}) \label{expand} ~.
\end{eqnarray}
We define a polynomial $B_{2l-4}$ of degree of $2l-4$ by
\bea
\sum_{i=1}^{l-1} \frac{\omega_i}{x (x^2-p_i^2)} = \frac{B_{2l-4}(x)}{x
H_{2(l-1)}(x)} ~. \label{b} \eea
with $H_{2(l-1)}(x)$ as in (\ref{HyF}). Then we have
\bea
\label{nec-1}
W_{cl}'(x) + \omega \Lambda ^{2N-2} x^{-1} = B_{2l-4}(x)
\sqrt{F_{4N-4l+2}(x) + \frac{\Lambda^{4N-4} x^2}{H_{2(l-1)}(x)^2}}
+ \CO (x^{-2}) ~. \eea
where $\omega = \mp \sum_{i=1}^l p_i^2 \omega_i$. Since the highest
order in $W_{cl}'(x)$ is $g_{n+1} x^{2 n +1}$, we see that $B_{2l-4}(x)$
should be of order $2n-2N+2l$. This shows that $l \geq N -n$ and for
$l=N-n$, $B_{2l-4}(x)$ becomes a constant equal to $g_{n+1}$. By squaring
(\ref{nec-1}), we obtain
\bea
g_{n+1}^2 F_{4n+2}(x) =   W'(x)^2 + 2g_{n+1}\omega \Lambda^{2N-2} x^{2n}
+ \CO(x^{2n-2}) ~.
\eea
We have thus showed $g_{n+1}^2 F_{4n+2} = W'^2 + f(x)$ (\ref{nec}),
and \bea f(x) =2g_{n+1}\omega \Lambda^{2N-2} x^{2n} + \CO(x^{2n-2}) ~.
\eea
To show that both sides of (\ref{nec}) are exactly the same, we would
need to consider lower order terms which depend nontrivially on $N_i$.
It is highly desirable to obtain this result exactly, which would
imply that the exact value of $\tau_{ij}(g_r,\Lambda)$, corresponding to
the $U(1)^n$ massless photons, found using the reduced $\CN =2$ curve
(\ref{red-curve}) evaluated in the $\CN=1$ supersymmetric vacua, is
consistent with that of (\ref{tauij}), found via the large $N$ duality
from the proposed geometric transition.

The result for $Sp$ is obtained in a similar way and we will not
give the details because they are identical to the ones of above.
The deformation will be of an identical form, this time the
adjoint field $\Phi$ being a symmetric 2-tensor and determines a
gauge symmetry breaking: \bea
 Sp(2 N) \rightarrow Sp(2 N_0) \times U(N_1) \times \cdots U(N_n) ~.
\eea In order to check the consistency, one compares the periods
obtained from the geometry which are stated in the section 2 and
the ones given by the reduced Seiberg--Witten curve discussed in
\cite{tera,ahn}. Finally, let us remark that another important
check of this duality for $SO/Sp$ gauge theories will be provided
by a detailed computation of the superpotential on both dual
theories.

~

{\bf Acknowledgements}

We are pleased to thank  Freddy Cachazo, Marcos Mari\~no, Javier
Mas, Carlos N\'u\~nez, C\'esar Seijas and Cumrun Vafa for discussions
and suggestions on the related matter. The work of JDE has been
supported by the Argentinian National Research Council (CONICET) and
by a Fundaci\'on Antorchas grant under project number 13671/1-55. The
work of KO is supported by NSF grant PHY-9970664.

\newpage


\end{document}